\documentclass[twocolumn]{aastex62}
\received{---}
\revised{----}
\accepted{--- }          

\usepackage{color}
\usepackage{graphicx}
\usepackage{amssymb}
\usepackage{array}
\usepackage{color}
\usepackage{pifont}
\usepackage{subfigure} 
\usepackage{csquotes}
\usepackage{natbib}
\bibpunct{(}{)}{;}{a}{,}{,}

\newcommand{\degree}{\ensuremath{^\circ}}

\def \sun {$_{\odot}$}

\shorttitle{Magnetic fields in L1521F}
\shortauthors{Soam et al.}

\begin{document}

\title{ First Sub-pc Sale Mapping of Magnetic Fields in the Vicinity of a Very Low Luminosity Object, L1521F-IRS}

\correspondingauthor{Archana Soam}
\email{archanasoam.bhu@gmail.com, asoam@usra.edu}

\author[0000-0002-6386-2906]{Archana Soam}
\affiliation{SOFIA Science Centre, USRA, NASA Ames Research Centre, MS-12, N232, Moffett Field, CA 94035, USA}
\affiliation{Korea Astronomy and Space Science Institute (KASI), 776 Daedeokdae-ro, Yuseong-gu, Daejeon 34055, Republic of Korea}

\author{Chang Won Lee} 
\affiliation{Korea Astronomy and Space Science Institute (KASI), 776 Daedeokdae-ro, Yuseong-gu, Daejeon 34055, Republic of Korea}
\affiliation{University of Science and Technology, Korea (UST), 217 Gajeong-ro, Yuseong-gu, Daejeon 34113, Republic of Korea}

\author{B-G Andersson}
\affiliation{SOFIA Science Centre, USRA, NASA Ames Research Centre, MS-12, N232, Moffett Field, CA 94035, USA}

\author{Maheswar G.}
\affiliation{Indian Institute of Astrophysics, Kormangala (IIA), Bangalore 560034, India}

\author{Mika Juvela}
\affiliation{Department of Physics, P.O.Box 64, FI-00014, University of Helsinki}

\author{Tie Liu}
\affiliation{Shanghai Astronomical Observatory, Chinese Academy of Sciences, 80 Nandan Road, Shanghai 200030, China}
\affiliation{Korea Astronomy and Space Science Institute (KASI), 776 Daedeokdae-ro, Yuseong-gu, Daejeon 34055, Republic of Korea}
\affiliation{East Asian Observatory, 660 N. A`oh\={o}k\={u} Place, University Park, Hilo, HI 96720, USA}

\author[0000-0003-2011-8172]{Gwanjeong Kim}
\affil{Nobeyama Radio Observatory, National Astronomical Observatory of Japan, National Institutes of Natural Sciences, Nobeyama, Minamimaki, Minamisaku, Nagano 384-1305, Japan}

\author{Ramprasad Rao}
\affiliation{Academia Sinica Institute of Astronomy and Astrophysics, P.O. Box 23-141, Taipei 10617, Taiwan}

\author{Eun Jung Chung}
\affiliation{Korea Astronomy and Space Science Institute (KASI), 776 Daedeokdae-ro, Yuseong-gu, Daejeon 34055, Republic of Korea}

\author{Woojin Kwon} 
\affiliation{Korea Astronomy and Space Science Institute (KASI), 776 Daedeokdae-ro, Yuseong-gu, Daejeon 34055, Republic of Korea}
\affiliation{University of Science and Technology, Korea (UST), 217 Gajeong-ro, Yuseong-gu, Daejeon 34113, Republic of Korea}

\author{Ekta S.}
\affiliation{Indian Institute of Astrophysics, Kormangala (IIA), Bangalore 560034, India}

\begin{abstract}
L1521F is found to be forming multiple cores and it is cited as an example of the densest core with an embedded VeLLO in a highly dynamical environment. We present the core-scale magnetic fields (B-fields) in the near vicinity of the VeLLO L1521F-IRS using submm polarization measurements at 850$~\mu$m using JCMT POL-2. This is the first attempt to use high-sensitivity observations to map the sub-parsec scale B-fields in a core with a VeLLO. The B-fields are ordered and very well connected to the parsec-scale field geometry seen in our earlier optical polarization observations and the large-scale structure seen in Planck dust polarization. The core scale B-field strength estimated using Davis-Chandrasekhar-Fermi relation is $\rm 330\pm100~\mu$G which is more than ten times of the value we obtained in the envelope (envelope in this paper is \enquote{core envelope}). This indicates that B-fields are getting stronger on smaller scales. The magnetic energies are found to be 1 to 2 orders of magnitude higher than non-thermal kinetic energies in the envelope and core. This suggests that magnetic fields are more important than turbulence in the energy budget of L1521F. The mass-to-flux ratio of 2.3$\pm$0.7 suggests that the core is magnetically-supercritical. The degree of polarization is steadily decreasing towards the denser part of the core with a power law slope of -0.86.


\end{abstract}




\section{Introduction} \label{sec:intro}

Magnetic fields (B-fields) are found to be one of the important drivers in the star formation process but they are not very well constrained by available observations. The influence of B-fields on various spatial scales and stages of star-formation is still unclear. Both magnetically dominated and turbulence dominated scenarios have been advocated to explain the fact that star formation is slow compared to free-fall times \citep{2005Natur.438..332K, 2006ApJ...641L.121T}. In the magnetically dominated scenario for isolated low mass star formation, the cores gradually condense out of a magnetically subcritical background cloud, through ambipolar diffusion \citep{1987ARA&A..25...23S, 1993prpl.conf..327M, 1999osps.conf..305M, 2003ApJ...599..363A}. In this process, the material, mediated by the magnetic field lines, settles into a disk-like morphology of a few thousand AU in size. This allows the cloud-scale magnetic fields to become parallel to the cloud minor axis. Turbulence-dominated scenarios will produce less orderly magnetic field configurations.

The B-fields are mapped using well established technique of polarization measurements. In the shorter wavelengths (i.e. optical and near infrared), the dichroism or selective extinction causes the observed polarization. The polarization position angles in optical wavelengths trace the plane-of-sky orientation of the ambient magnetic field at the periphery of molecular clouds (with $\rm A_{V}\approx$ 1$-$2 mag) \citep{1995ApJ...448..748G, 1996ASPC...97..325G}. The minor axes of dust grains align with the orientation of magnetic fields \citep{2008MNRAS.388..117H}. Polarization observations at submillimetre/millimetre wavelengths are used to trace the field lines in more dense regions of the cloud \citep{2009MNRAS.398..394W}. The thermal emission from the dust grains is stronger along the major axis, producing a polarization direction perpendicular to the magnetic fields in the denser core regions ($\rm n_{H_{2}} \sim 10^{5}-10^{6}~cm^{-3}$) observable even with $\rm A_{V} >$50 mag. The line-of-sight component of the magnetic field is measured using Zeeman observations \citep{1993ApJ...407..175C}.

L1521F \citep[MC27;][]{1997A&A...324..203C, 1999PASJ...51..257O, 2001ApJS..136..703L} is located in the Taurus star- forming region at a distance of 140 pc \citep{2005ApJ...619L.179L, 2007ApJ...671.1813T}. This core contains a very low luminosity object (VeLLO) L1521F-IRS with bolometric luminosty of L = 0.05$~\rm L_{\odot}$ \citep{2006ApJ...637..811T, 2006ApJ...649L..37B}. The source was discovered by \textit{Spitzer Space Telescope} (SST) \citep{2006ApJ...649L..37B} but could not be detected in IRAS mission \citep{1986ApJ...307..337B, 1989ApJS...71...89B, 1997A&A...324..203C} due to its low luminosity. The core appears to be isolated and associated with strong central condensation in 160 $\mu$m \textit{Spitzer} image \citep{2007MNRAS.375..843K} but recent ALMA observations using high density tracers reveal that this is a possible site of multiple star formation \citep{2014ApJ...789L...4T}.  The embedded VeLLO L1521F-IRS is found to be associated with a compact but poorly collimated molecular outflow in CO(J=2-1) line observations \citep{2013ApJ...774...20T}. Moreover, \citet{2014ApJ...789L...4T} found a very compact bipolar outflow centered at L1521F-IRS using HCO$^{+}$ (J=3-2) observations from ALMA. This core is considered to be at a very early stage of star formation based on recent ALMA observations of dust continuum emission and molecular rotational lines by \citet{2016ApJ...826...26T}. In a latest study on CO outflow survey of 68 VeLLOs, \citet{2019ApJS..240...18K} suggest that L1521F-IRS is a potential proto-brown dwarf candidate. They reported a largest identified sample of 15 proto-brown dwarfs among 68 VeLLOs. Their segregation is based on the smaller envelope masses and lower accretion rates estimated from CO, $\rm ^{13}CO$, and $\rm C^{18}O$ molecular line observations.

VeLLOs (L$_{int}\lesssim 0.1\,L_{\odot}$) have been detected based on the data from the SST. The VeLLOs are interesting sources as their luminosity is an order of magnitude lower than the accretion luminosity $L_{acc}$ $\sim 1.6\,L_{\odot}$ expected for a 0.08\,$M_{\odot}$ protostar with an accretion rate of $\sim 10^{-6}\,M_{\odot}\,yr^{-1}$ \citep{1987IAUS..115..417S} and 3$R_{\odot}$ stellar radius. It has been speculated that these sources are either progenitors of proto-brown dwarfs \citep[e.g. ][]{2013ApJ...777...50L, 2016ApJS..222....7L} or very low mass protostars.

The outflows from the protostars are thought to influence their surrounding environment by generating turbulence which could scramble a relatively weak magnetic field in their vicinity. This could disturb any initial alignment between the core and the envelope magnetic fields. On the other hand, the estimated outflow parameters suggest that VeLLOs have the most compact, lowest mass, and the least energetic outflows compared to known Class 0/I outflows from low-mass stars \citep[e.g. ][]{2002A&A...393..927B, 2004A&A...426..503W, 2006ApJ...649L..37B, 2011ApJ...743..201P}. We therefore expect the outflows from these sources to have the least significant effect on their surroundings. This would enable the regions to preserve the initial condition that may have existed prior to the initiation of star formation showing primordial magnetic fields.

Here we present the first high-sensitivity submm polarization observations made towards the L1521F core harboring a VeLLO with the Submillimetre Common-User Bolometer Array 2 (SCUBA-2) camera with the POL-2 polarimeter commissioned at James Clarke Maxwell Telescope (JCMT). The mapping of sub-pc scale B-fields towards a low mass star-forming core (with a VeLLO or proto-brown dwarf candidate) in a highly dynamical environment is presented for the first time in the present study. The line-of-sight magnetic field towards this source is already studied by \citet{2010ApJ...725..466C}. The molecular line observations towards this region using the Heterodyne Array Receiver Program \citep[HARP;][]{2016A&A...586A..44C} and Atacama Large millimeter/submillimeter Array \citep[ALMA;][]{2014ApJ...789L...4T, 2016ApJ...826...26T} are already available to understand the kinematics of the core.

The paper is organized as follows: in Section 2, we describe the observations and data reduction; in Section 3, we give initial results; in Section 4 and 5, we discuss and summarize our results.

\section{Data acquisition and reduction techniques} \label{sec:obs}

We observed L1521F in 850 $\mu$m polarized emission with SCUBA-2 \citep{2013MNRAS.430.2513H} in conjunction with POL-2 \citep{2016SPIE.9914E..03F, 2019inpreparation} in the nights of Nov. 11 to 14, 2017 under the project code M17BP070 (PI: Soam A.) at JCMT.  L1521F was mapped with 18 observations, with an average integration time of $\sim0.55$ hours per observation, in good weather (0.05 $<\tau_{225}<$ 0.08, where $\tau_{225}$ is atmospheric opacity at 225\,GHz). A POL-2 daisy \citep{2016SPIE.9914E..03F} scan pattern was used for mapping the core producing a uniform, high signal-to-noise coverage over the central 3${\arcmin}$ of the map. This pattern is similar to SCUBA-2 CV daisy scan pattern \citep{2013MNRAS.430.2513H} but modified to have a slower scan speed (8${\arcsec}$/s compared to 155${\arcsec}$/s) to obtain sufficient on-sky data for good Stokes Q and U values. A fully-sampled mapping was done in a 12${\arcmin}$ diameter, circular region with an effective resolution of 14.1${\arcsec}$. The frequency of wave-plate rotation was 2 Hz.

For reducing the data, we used the {\tt\string pol2map} python script in the Starlink \citep{2014ASPC..485..391C}\footnote{\software{Starlink \citep{2014ASPC..485..391C}, SMURF \citep{2013ascl.soft10007J}} used in POL-2 data reduction is currently supported by the East Asian Observatory.} {\tt\string SMURF} \citep{2013ascl.soft10007J} package. Starlink tasks such as {\tt\string calcqu} and {\tt makemap} \citep{2013MNRAS.430.2545C} were used in the reduction process. The detailed procedure to reduce the data and to produce polarization catalog is explained in \citet{2018ApJ...859....4K} and \citet{2018ApJ...861...65S}. We adopted slightly different reduction procedure by using an additional parameter {\tt\string skyloop} in {\tt\string pol2map} setting pixel size as 12${\arcsec}$. See \citet{2015MNRAS.454.2557M} and \citet{2018ApJ...859....4K} for a more detailed description of the SCUBA-2 and POL-2 data reduction processes, respectively. The vectors are de-biased \citep{2018ApJ...859....4K} to remove the effect of statistical biasing in low signal-to-noise-ratio (SNR) regions. Polarized values are obtained by combining Q and U maps where polarized intensity exceeds 2 times its standard deviation. For the analysing the data, we have used only those detections where S/N$>2$ in polarized intensity and polarization fraction. All position angles are measured from north increasing towards east.  Magnetic field orientation is derived by rotating the polarization angles by 90${\degree}$.

\begin{figure*}
\begin{center}
\resizebox{18.0cm}{14.0cm}{\includegraphics{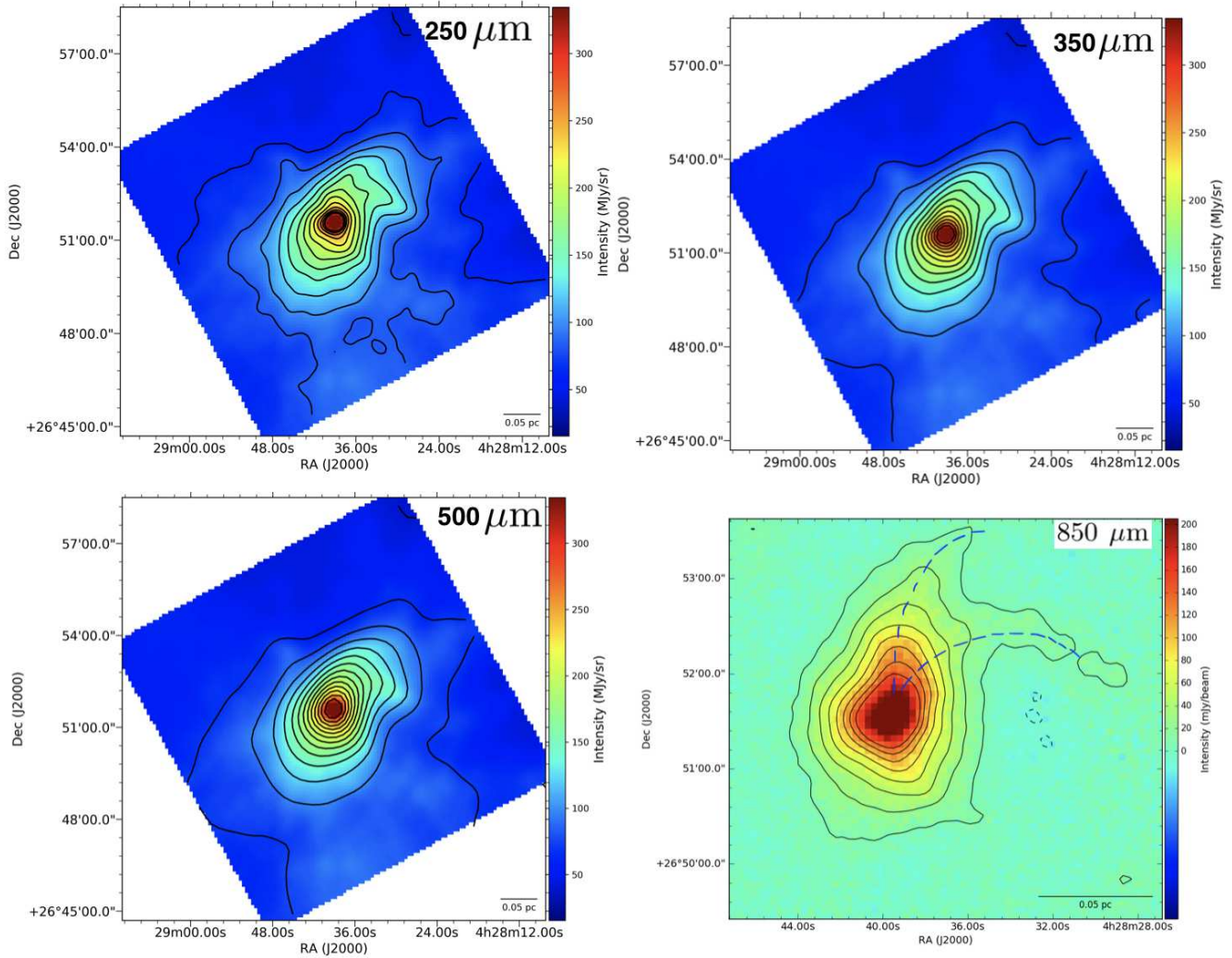}}
\caption{The dust emission maps of L1521F seen at 250, 350, and 500$~\mu$m wavelengths by \textit{Herschel}/SPIRE and at 850$~\mu$m by JCMT/SCUBA-2. Contours of  850\,$\mu$m intensity are shown with levels 8, 10, 45, 80, 105, 140\,$\rm mJy\,beam^{-1}$. The two interesting features seen at 850\,$\mu$m are indicated with blue dashed lines.}\label{Fig:stokesI}
\end{center}
\end{figure*}

\begin{table}
\centering
\caption{Results of JCMT/POL-2 observations towards L1521F at 850\,$\mu$m wavelength.}\label{tab:pol_res}
\scriptsize
\begin{tabular}{lllllc}\hline
 Id  & $\alpha$ (J2000)  & $\delta$ (J2000)  & I$\pm \sigma_I$ & P $\pm$ $\sigma_P$ & $\theta$ $\pm$ $\sigma_{\theta}$  \\ 
     &($\degree$)&($\degree$)&(mJy/beam) &(\%) &($\degree$) \\\hline  
\hline
1   &67.164 &  26.848 & 136.6 $\pm$ 3.1 & 12.6 $\pm$  3.6   &    -75.8 $\pm$ 10.3 \\          
2   &67.160 &  26.848 & 105.8 $\pm$ 1.8 & 8.3  $\pm$  3.7   &    -89.4 $\pm$ 9.5  \\         
3   &67.156 &  26.848 & 54.5  $\pm$ 2.2 & 14.8 $\pm$  6.4   &    -82.7 $\pm$ 7.3  \\         
4   &67.164 &  26.851 & 100.4 $\pm$ 5.3 & 5.5  $\pm$  1.9   &    -64.3 $\pm$ 9.2  \\         
5   &67.160 &  26.851 & 81.7  $\pm$ 2.7 & 11.0 $\pm$  1.8   &    -74.6 $\pm$ 4.9  \\         
6   &67.156 &  26.851 & 53.1  $\pm$ 1.3 & 14.1 $\pm$  4.6   &    -82.9 $\pm$ 12.4 \\          
7   &67.171 &  26.855 & 128.3 $\pm$ 1.8 & 3.1  $\pm$  1.4   &    -44.4 $\pm$ 11.9 \\          
8   &67.164 &  26.855 & 176.3 $\pm$ 2.7 & 2.9  $\pm$  0.8   &    -61.8 $\pm$ 7.8  \\         
9   &67.160 &  26.855 & 105.3 $\pm$ 1.9 & 3.6  $\pm$  1.3   &    -87.8 $\pm$ 9.8  \\        
10  &67.153 &  26.858 & 42.5  $\pm$ 1.0 & 15.8 $\pm$  7.8   &    -52.5 $\pm$ 12.1 \\          
11  &67.175 &  26.861 & 95.3  $\pm$ 2.2 & 9.5  $\pm$  4.4   &    -39.8 $\pm$ 11.0 \\          
12  &67.171 &  26.861 & 143.0 $\pm$ 1.4 & 4.7  $\pm$  1.6   &    -66.0 $\pm$ 9.2  \\           
13  &67.175 &  26.865 & 55.1  $\pm$ 2.1 & 23.8 $\pm$  9.9   &    -51.6 $\pm$ 9.8  \\         
14  &67.171 &  26.865 & 92.4  $\pm$ 2.5 & 8.7  $\pm$  3.4   &    -59.8 $\pm$ 9.3  \\         
15  &67.168 &  26.865 & 115.0 $\pm$ 1.1 & 3.8  $\pm$  1.2   &    -32.5 $\pm$ 9.6  \\         
16  &67.171 &  26.868 & 72.6  $\pm$ 1.5 & 10.8 $\pm$  4.8   &    -63.8 $\pm$ 10.5 \\          
17  &67.168 &  26.868 & 98.5  $\pm$ 2.7 & 5.2  $\pm$  2.1   &    -22.0 $\pm$ 10.6 \\          
18  &67.164 &  26.868 & 123.4 $\pm$ 2.1 & 3.3  $\pm$  1.6   &    -42.9 $\pm$ 12.1 \\
\hline
\end{tabular}\\

$\theta$ is the polarization position angle before rotating by 90$\degree$  \\
\end{table}


\section{Results and Discussion}

\subsection{Structure and kinematics of L1521F}\label{Results}
Figure \ref{Fig:stokesI} shows the dust continuum emission maps of L1521F observed from \textit{Herschel}/SPIRE at 250, 350, and 500$~\mu$m wavelengths and JCMT/SCUBA-2 observations at 850$~\mu$m wavelength (this work). SCUBA-2 has a spatial resolution of 14${\arcsec}$ i.e. $\sim$ 2000 AU spatial scale at a distance of L1521F i.e. 140 pc.  The peak values of total and polarized intensities are found to be $\sim$ 200\,$\rm mJy\,beam^{-1}$ and $\sim$ 15\,$\rm mJy\,beam^{-1}$, respectively. The rms noise of the background region in the Stokes I map is measured to be $\rm \sim 2.5 ~mJy~beam^{-1}$. This value was estimated by selecting a region with relatively constant signal $\sim1{\arcmin}$ away from the center of Stoles I map. This regions is relatively flat, moderately unpolarized, low in emission, and away from the brightest region in the I map. The standard deviation of the measured flux density distribution in that region is considered as the rms of the Stokes I map.

The L1521F core seems to be elongated in north and south.  The 850$~\mu$m observation indicates two curved emission features that seem to be coming out of the core in northward direction and one of them is bending westward. These features are indicated by dashed lines in panel (d) of the figure \ref{Fig:stokesI}. Interestingly, the similar core structure can be noted in \textit{Herschel}/SPIRE images shown in panels (a), (b), and (c) of figure \ref{Fig:stokesI} at different wavelengths. The result of JCMT/POL-2 polarization observations are given in Table \ref{tab:pol_res}.

\citet{1999sf99.proc..177L} classified L1521F core as an infall candidate based on their high resolution observations of CS (J=2-1) which is supported by \citet{1999PASJ...51..257O} from their $\rm HCO^{+}$ (J=3-2) and (J=4-3) observations. The CS (J=2-1) and $\rm N_{2}H^{+}$ (J=1-0) maps of L1521F shown by \citet{2001ApJS..136..703L}, however, indicate an extended red asymmetry over the core region. The formation of L1521F-IRS can be correlated with these motions seen by \citet{2001ApJS..136..703L}.

L1521F is found to be embedded in a very interesting and dynamic environment seen by the ALMA $\rm HCO^{+}$ (J=3-2) observations by \citet{2014ApJ...789L...4T}. The compact bipolar outflow seen in their $\rm HCO^{+}$ (J=3-2) map at $\sim$500 AU scale shows some dynamical interaction with the surrounding gas. There are two high density cores detected in both dust continuum emission and $\rm H^{13}CO^{+}$ (J=3-2). The observed features seen in the $\rm HCO^{+}$ (J=3-2) and $\rm H^{13}CO^{+}$ (J=3-2) lines by \citet{2014ApJ...789L...4T} suggest that L1521F is probably embedded in a site where multiple stars are being formed. \citet{2016ApJ...826...26T} presented ALMA observations of dust continuum emission and molecular rotational lines towards L1521F. The two starless cores in the close vicinity of L1521F are found and complex gas structure is seen in $\rm ^{12}CO$(J=3-2) and $\rm HCO^{+}$(J=3-2) observations. An arc-like feature and a few other core features connected to L1521F are noticed in $\rm HCO^{+}$ (J=3-2) emission. The length of the arc-like structure was found to be  $\sim$2000 AU. They considered this arc-like structure as a possible result of dynamical interaction between the small dense cores and the surrounding gas on $\sim$2000 AU scale. 


The two diffused features coming out of the L1521F core in 850$~\mu$m and Herschel images (see figure \ref{Fig:stokesI}), are neither quite related to the observed magnetic field morphology not outflow cavity in the L1521F core (see figure \ref{Fig:MF}). This could be one of the several such structures seen toward the Taurus molecular clouds. Several of these structures show similar east-west patterns, for example, the structures seen in the direction of other L1521 cores (A, B, E etc.) suggesting some sort of an external influence. Recently, \citet{2019A&A...623A..16S} showed a possible scenario where the gas motion toward B213 filament is due to its interaction with Per OB2 association. Authors propose that B211/B213 filament was initially formed by large-scale compression of HI gas by Per OB2 association and then it is growing due to gravitational accretion of ambient cloud. L1521F is one of the several cores embedded in Taurus molecular cloud and may be undergoing in the process of gravitational accretion of surrounding material.


\subsection{Magnetic field morphology}\label{MF}

\begin{figure*}
\centering
\resizebox{18cm}{14cm}{\includegraphics{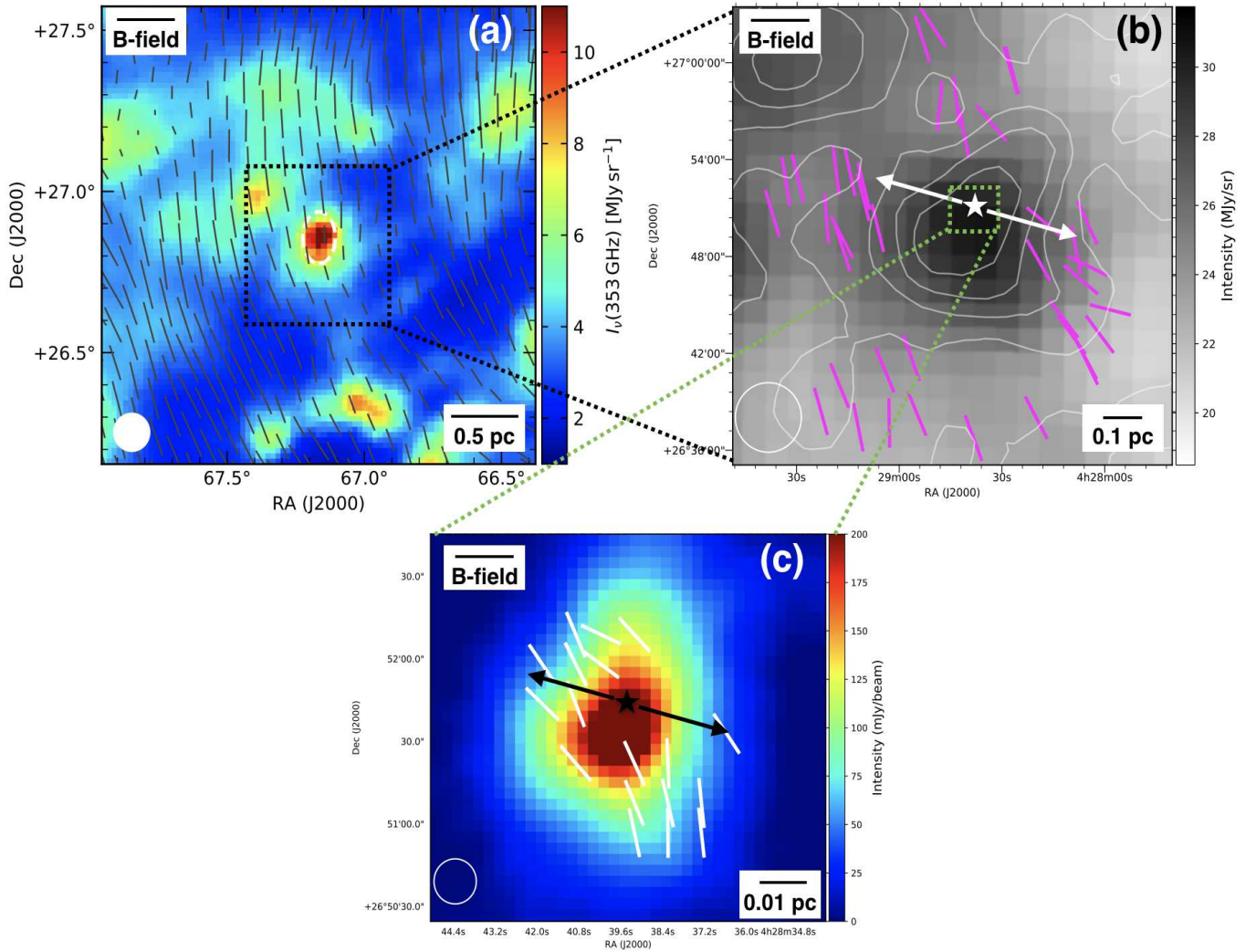}}
\caption{{\bf Panel (a):} Morphology of B-fields obtained from \textit{Planck} 850$~\mu$m dust polarization observations. The location of L1521F is shown with dashed ellipse in the center. {\bf Panel (b):} The B-fields mapped with optical R-band (0.63$~\mu$m) observations by \citet{2015A&A...573A..34S} plotted on IRAS 100$~\mu$m with the contours of Dobashi extinction map of L1521F \citep{2005PASJ...57S...1D}. The position of VeLLO and associated CO bipolar outflows \citep{2013ApJ...774...20T} are shown with star symbol and double headed arrow. {\bf Panel (c):} The B-field morphology obtained from 850$~\mu$m dust polarization observations of L1521F core shown on 850$~\mu$m dust continuum map. The lengths of line-segments are normalized and independent of fraction of polarization. The beam sizes in all the frames are shown with filled and open circles.} \label{Fig:MF}
\end{figure*}

\begin{figure}
\resizebox{9.5cm}{8.5cm}{\includegraphics{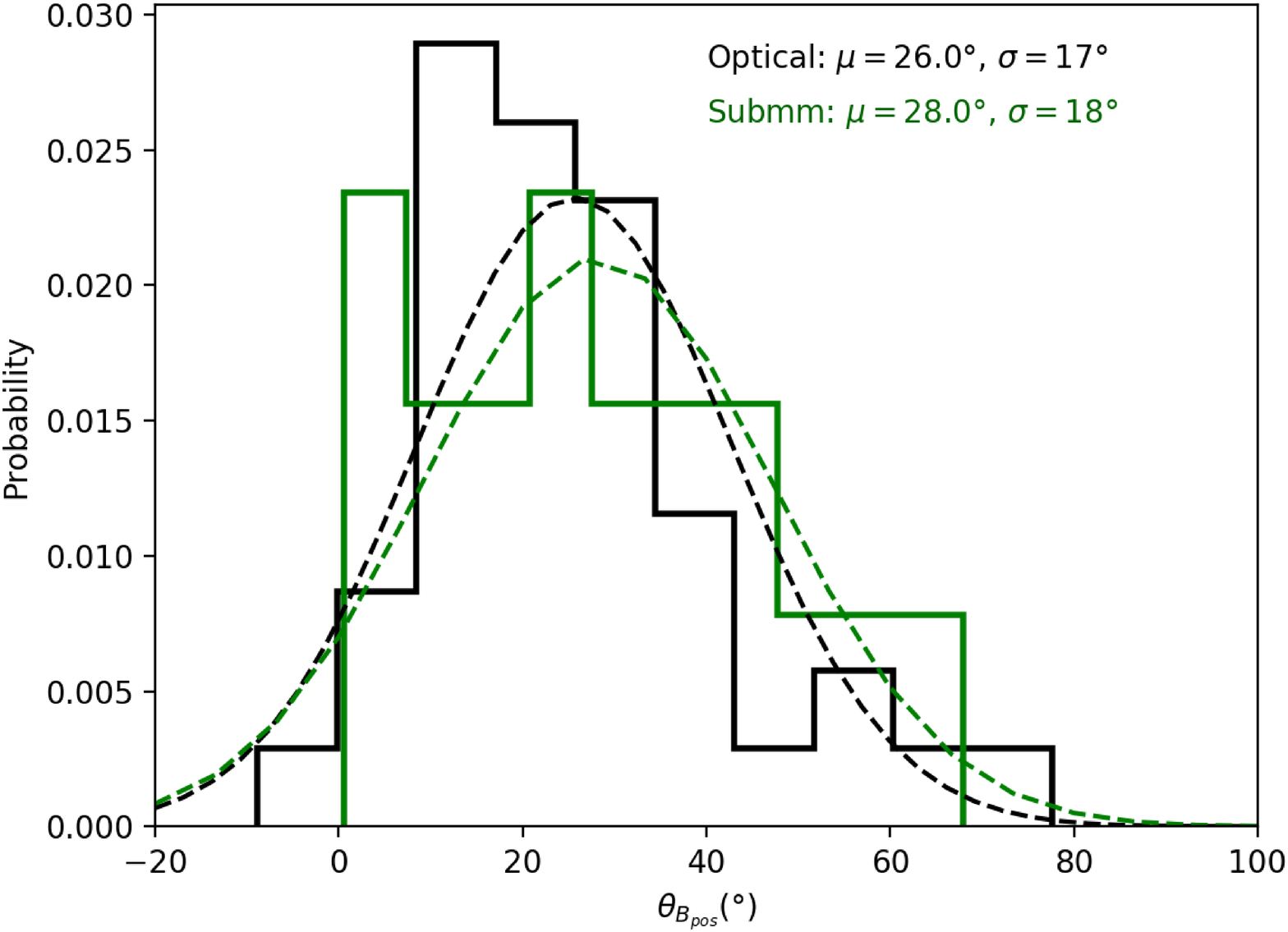}}\\
\resizebox{9.5cm}{8.5cm}{\includegraphics{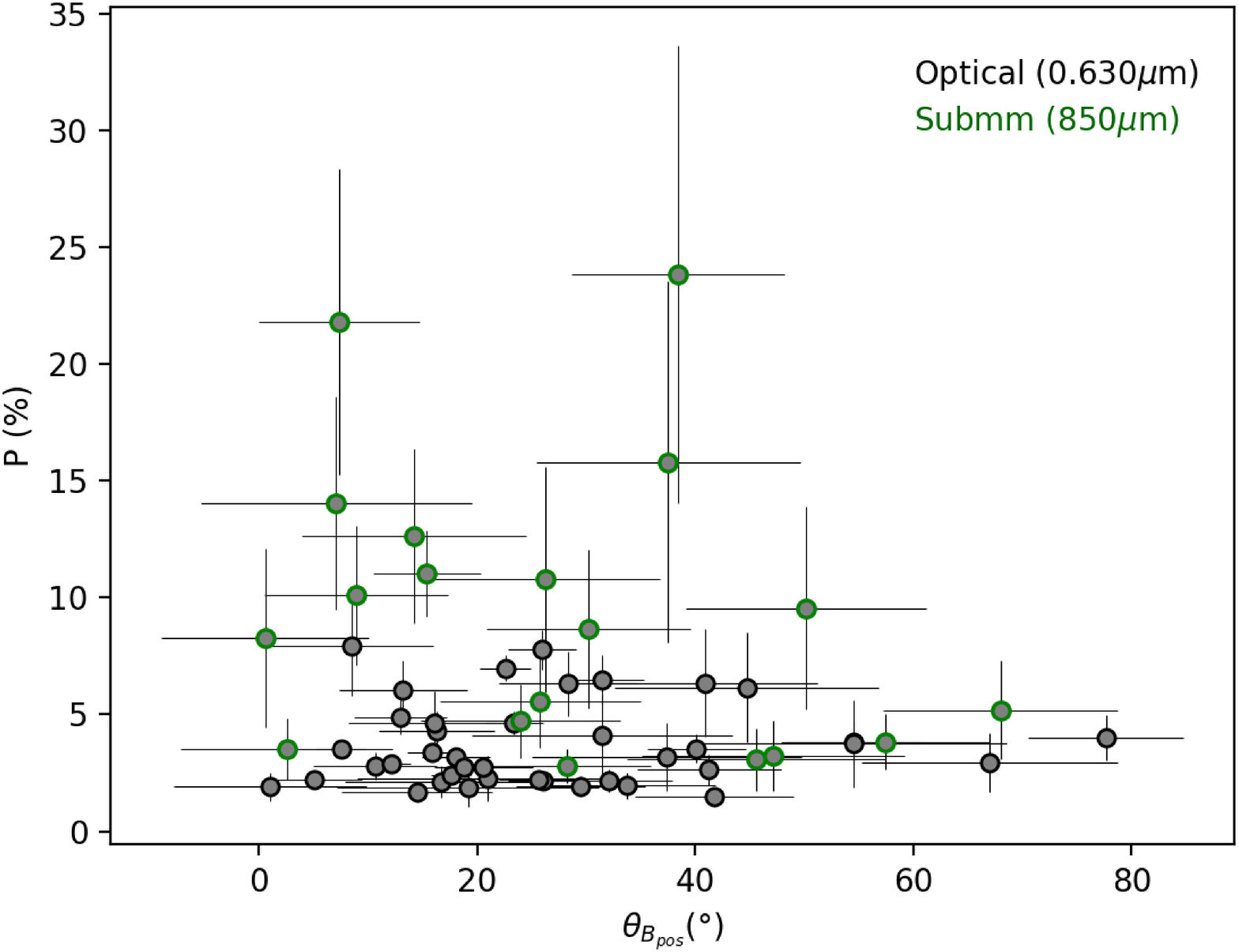}}
\caption{{\bf Upper panel:} The Gaussian fitted histograms of plane-of-sky B-field position angles measured at optical (black) and submm (green) wavelength towards L1521F. {\bf Lower panel:} Distribution of percentage of polarization with B-field position angles in the two samples.}\label{Fig:opt_sub_hist}
\end{figure}

Figure \ref{Fig:MF} shows the B-field morphology in L1521F obtained from Planck 850$~\mu$m, optical R-band (0.63$~\mu$m) and JCMT/POL-2 850$~\mu$m observations. The parsec scale B-field geometries are shown in panels (a) and (b), and the core field geometry at sub-parsec scale is zoomed-in in panel (c). The location of L1521F-IRS and the associated bipolar outflows is shown with star and double-headed arrow symbols in panels (b) and (c). 
Figure \ref{Fig:opt_sub_hist} shows the Gaussian fitted histogram to the distribution of plane-of-sky B-field position angles measured in optical and submm observations. The mean and standard deviation values are found to be similar in both the distributions. The lower panel of the same figure shows the distributions of polarization fraction and position angle values detected in envelope and core of L1521F. The maximum amount of polarization seen in diffuse envelope and dense core of L1521F are found to $\sim7\pm2$\% and $\sim24\pm10$\%, respectively. However, the pattern of distribution seems similar in both the cases.

Dust polarization in Planck 850$~\mu$m observations is used to investigate the B-field morphology at a scale of 5$\arcmin$ \citep{2016A&A...594A...1P}. The image in panel (a) of figure \ref{Fig:MF} is smoothed down to the 7$\arcmin$ resolution to ensure good SNR data. The vectors are drawn at 3.5$\arcmin$ (half-resolution) steps. The figure shows the large scale magnetic fields towards L1521F. The location of L1521F is shown by the white ellipse indicating the area covered by SCUBA-2 observations. The large-scale field seems to be running from north-east to south-west. However, at the location of L1521F, the fields lines are mainly seen in north-south direction. A smooth bending in the field lines can be seen from south-west to north direction. The pinching of field lines is perceptible in the region east to L1521F. This region also shows a higher degree of polarization than seen in L1521F. Similar trend of field lines was observed by \citet{2015A&A...573A..34S} when they plotted the optical polarization vectors with the Heiles catalog \citep{2000AJ....119..923H} data selected within 1$\degree$ radius around L1521F (see figure 8 of \citealt{2015A&A...573A..34S}).

The field geometry in L1521F envelope seen from optical observations are zoomed in panel (b) of figure \ref{Fig:MF} \citep{2015A&A...573A..34S}. The mean orientation of field lines is from north-east to south-west directions which seems consistent to that seen in Planck observations. The field lines appear organized but not aligned with the outflow direction.  

The B-field geometry is investigated in L1521F from POL-2 850\,$\mu$m observations by further zooming-in (panel (c) of figure \ref{Fig:MF}). Interestingly, the north-east to south-west component is still seen in the map. The difference between the outflow direction and the magnetic field orientation in the envelope is $\sim 50 \degree$ \citep{2015A&A...573A..34S}.  The core scale B-fields are also misaligned with an offset of $\sim 48 \degree$ between outflows and mean magnetic field orientation. However, field lines are more aligned with outflows if the north-eastern component only is taken into account.

The maps seen at three different scales suggest that field lines on envelope and core scales are connected and the global B-field orientation is from north-east to south-west. This connection is also depicted in the histograms of position angles plotted in figure \ref{Fig:opt_sub_hist}. Both populations peak between $\sim 0-40 \degree$. The correlation of optical and submm observations can further suggest that the core is embedded in strong magnetic field environment.

\begin{figure}
\resizebox{8.0cm}{8.0cm}{\includegraphics{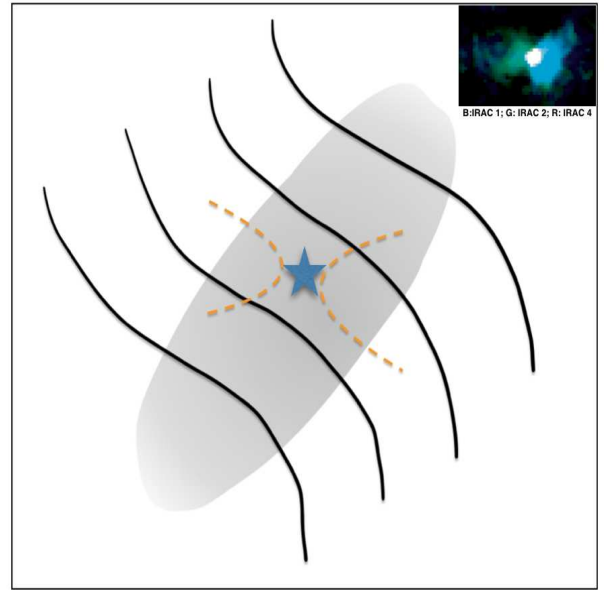}}
\caption{Cartoon showing the field lines, location of L1521F-IRS and bipolar outflow cavities in L1521F. The inset in upper right corner shows the \textit{Spitzer} color composite image obtained from \citet{2006ApJ...649L..37B}.}\label{Fig:cartoon}
\end{figure}

An interesting feature of bending in field lines can be seen in the core scale B-field geometry. This is further demonstrated in figure \ref{Fig:cartoon} where elliptical core and the field lines are shown. The location of embedded source in the core is shown with a star symbol. The associated bipolar outflow cavities are indicated with dashed parabolas based on the \textit{Spitzer} observations of this source by \citet{2006ApJ...649L..37B} shown in color-composite image as inset in the upper-right corner of this figure. The field lines in this cartoon seem to be result of north-east to south-west orientation of Planck and optical polarization observations which gets bent in core seen as submm wavelength.


\subsection{Strength of magnetic fields}\label{strength}
We used  Davis-Chandrasekhar-Fermi method (hereafter DCF method; \citealt{1951PhRv...81..890D, 1953ApJ...118..113C}) for field strength estimation in L1521F core. The modified DCF relation \citep{2004ApJ...600..279C} for measuring the plane-of-sky B-field ($\rm B_{pos}$) strength is:

\begin{equation}\label{eq:DCF}
B_{pos} = Q^{\prime} \sqrt{4 \pi \rho}\frac{\sigma_{v}}{\delta \theta} \approx 9.3\sqrt{n({\rm H}_{2})}\frac{\Delta v}{\delta \theta}\,\mu{\rm G},
\end{equation}

where $Q^{\prime}$ accounts for variation in field strength on scales smaller than beam and is taken as a factor of order unity \citep{2004ApJ...600..279C}. The value of $Q^{\prime}$ here is used as 0.5 \citep{2001ApJ...546..980O}. The $\rho = \mu_{g}m_{H}n({\rm H}_{2})$ is the gas density where $\mu_{g}$ = 2.8 is the mean molecular weight of the gas \citep{2008A&A...487..993K}, $m_{H}$ is the mass of an hydrogen atom, and $n({\rm H}_{2})$ is the number density of molecular hydrogen in cm$^{-3}$. In eq. \ref{eq:DCF}, $\Delta v$  = $\sigma_{v}\sqrt{8\ln 2}$ is FWHM in km\,s$^{-1}$ where $\sigma_{v}$ is the average line-of-sight non-thermal velocity dispersion, $\delta \theta$ is corrected dispersion in position angle in degrees.  The corrected dispersion in position angle is obtained by adopting the procedure explained by \citet{2001ApJ...561..864L} and \citet{2010ApJ...723..146F} where correction is done in quadrature by using $\delta \theta = ({\sigma_{std}}^{2} - {\langle\sigma_{\theta}\rangle}^{2}$)\textsuperscript{1/2}, where $\sigma_{std}$ is the standard deviation in the distribution of position angle and the mean uncertainty $\langle\sigma_{\theta}\rangle$ was estimated from $\langle\sigma_{\theta}\rangle = \Sigma\sigma_{\theta i}/N$, with $\sigma_{\theta i}$ as the uncertainty in  $i^{th}$ polarization angle\footnote{The uncertainty in the position angles is calculated by error propagation in the expression of polarization angle $\theta$, which gives, $\sigma_{\theta} = 0.5\times\sigma_{P}/P$ in radians, or $\sigma_{\theta} = 28.65\degree\times\sigma_{P}/P$ \citep[see;] []{Serkowski1974} in degrees.} and N as number of position angles.

The corrected value of dispersion in position angle $\sigma_{\theta}$ is found to be $\sim15\pm2\degree$ where uncertainty is measured as standard deviation in the distribution of measured uncertainties in the position angles. The turbulence in the DCF relation is measured by the dispersion in the line-of-sight velocity. \citet{2016A&A...586A..44C} has done a detailed investigation of kinematics in cloud condensations towards Taurus and Perseus star-forming regions using JCMT HARP observations of $\rm HCO^{+} (J=4-3)$ and $\rm C^{18}O (J=3-2)$ gas tracers.  But the $\rm C^{18}O (J=3-2)$ line profiles are affected by the motions in the core envelope. Therefore we looked at the survey towards starless cores presented by  \citet{2001ApJS..136..703L} using $\rm N_{2}H^{+}$ (J=1-0) lines. We have adopted the FWHM value of $\rm N_{2}H^{+}$ (J=1-0) line ($\Delta v_{N_{2}H^{+}}$ = 0.37$\pm$0.02 $\rm km s^{-1}$) observed towards L1521F core. This core was originally noticed as a dense condensation with a high central density of $\rm \sim 10^{6}~ cm^{-3}$ \citep{1999PASJ...51..257O}. \citet{2005MNRAS.360.1506K}  reported a volume density of L1521F core as $\rm 2\times10^{6}~cm^{-3}$ with $\sim$50\% uncertainty using JCMT/SCUBA mapping. We used the above values of standard deviation in polarization angles, dispersion in $\rm N_{2}H^{+}$ (J=1-0) line velocity and volume density in DCF relation and estimated the B-field strengths as $\rm \sim330\pm100~\mu$G in the core. 

Plane-of-the-sky magnetic field strength estimation in the envelope of L1521F ($\rm \sim 25~\mu$G) by \citet{2015A&A...573A..34S} is found to be more than ten times lower than the value we obtained at the core-scale. This suggests a strengthening of B-fields in the sub-pc scale from parsec scale. This may be caused by the dragging of field lines from the envelope to the core region. The line-of-sight B-field strength towards this core has been found to be -1.4$\pm$4.0$~\mu$G by \citet{2010ApJ...725..466C} using OH Zeeman observations from Arecibo telescope. The plane-of-sky and line-of-sight magnetic field strengths suggest that the average B-field strength in the core is of the order of $\sim330~\mu$G.

\subsection{Mass-to-flux ratio}\label{planck}

The mass-to-flux ratio estimation helps in testing the relative importance of gravity and magnetic fields in the molecular clouds. This parameter is represented by $\lambda$ \citep{2004Ap&SS.292..225C}. The core stability can be tested using the observed B-field strength and column density values. The $\rm H_{2}$ column density of L1521F is found to be $\rm 1 \times 10^{23}~cm^{-2}$ by \citet{2005MNRAS.360.1506K} from JCMT/SCUBA observations. We adopted this value of column density and the average magnetic field strength as $\rm \sim330\pm100~\mu$G in L1521F core. The value of $\lambda$ is estimated using the the relation given by \citet{2004Ap&SS.292..225C},

\begin{equation}\label{lambda}
\mathrm{\lambda = 7.6 \times 10^{-21} \frac{N(H_{2})}{B_{pos}}} ,
\end{equation}
 Where $\rm N(H_{2})$ is the molecular hydrogen column density in $\mathrm{cm^{-2}}$ and $\mathrm{B_{pos}}$ is the plane-of-sky field strength in $\mu$G. For the considered values of $\rm B_{pos}$ and column density, we find $\lambda=$2.3$\pm$0.7 where the error is included from the uncertainty in B-field strength measurement only. Since the plane-of-sky magnetic field is dominating towards L1521F, we are not applying any geometrical correction on the measured value of $\lambda$. This suggest that the core is super-critical which is consistent with the asymmetric line-profiles seen by \citet{2001ApJS..136..703L} predicting global infall motions in the core.
 
 \subsection{B-fields and turbulence from envelope to core}\label{planck}
 
 A simple comparison of magnetic field strength and turbulence from parsec to subparsec scales can be done in L1521F using optical and submm polarization results and line observations using $\rm CO$ (J=1-0) in diffuse envelope \citep{2019ApJS..240...18K} and $\rm N_{2}H^{+}$ (J=1-0) in dense core region \citep{2001ApJS..136..703L}. Since plane-of-the-sky component of magnetic fields is much stronger than line-of-sight field strength, we assume the total field strength ($\rm B_{tot}$) same as the measured value of plane-of-the-sky B-field strength. 

 The total magnetic field energy ($\rm E_{mag}$) can be calculated as
 
 \begin{equation}
 E_{mag} = \frac{B_{tot}^{2}V}{2\mu_0},
 \end{equation}
 where V is the core volume ($\rm =4/3\times\pi r^{3}$) with r as radius of core. $\mu_0$ is the permeability of vacuum. We used radius of envelope as the extent up to which optical polarization measurements are done. The core radius is estimated from the $850~\mu$m continuum emission.  

The non-thermal kinetic energy ($\rm E_{NT,kin}$) of the envelope and core can be calculated as
 \begin{equation}
 E_{NT,kin} = \frac{3M\sigma_{v}^2}{2},
 \end{equation}
 where M is mass of the core and $\sigma_{v}$ is the average one-dimensional line-of-sight non-thermal velocity dispersion. We adopted the envelope mass from \citet{2016ApJS..225...26K}, measured from available \textit{Herschel}/SPIRE 250 or 500 $\mu$m fluxes. We estimated the core mass from observed 850$~\mu$m emission using the following relation \citep{1983QJRAS..24..267H}
 
 \begin{equation}
 M = \frac{S_{\nu}Dd^{2}}{\kappa_{\nu}B_{\nu}(T)},
  \end{equation}
 where $S_{\nu}$ is the 850$~\mu$m flux density, D is the dust-to-gas mass ratio assumed as 0.01, d is the distance of the core (i.e 140 pc), $\kappa_\nu$ is the dust opacity which is adopted as 1.85$\rm~cm^{2}g^{-1}$ from \citet{1994A&A...291..943O}, and $B_{\nu}$(T) is the Planck function. We assumed a core temperature of  9 K from the SCUBA observations of the L1521F at 850$~\mu$m \citep{2005MNRAS.360.1506K}.

 The values of $\rm B_{tot}$, $\rm E_{mag}$, and $\rm E_{NT,kin}$ for L1521F envelope and core are given in Table \ref{tab:energies}. The magnetic energies in the envelope and core are 1-2 orders of magnitude higher than the non-thermal kinetic energies of these regions suggesting that magnetic fields are more important than turbulence and contributing more to the energy budget of L1521F.


\begin{table}
\centering
\caption{Values calculated in L1521F envelope and core regions.}\label{tab:energies}
\scriptsize
\begin{tabular}{llllc}\hline
Region  & $\rm B_{tot}$ & $\rm E_{mag}$ & $\rm E_{NT,kin}$  \\
        & ($\mu$G) & (J) & (J) \\\hline
Envelope &25 &$\rm 1.3\times 10^{38}$  & $\rm 1.9\times 10^{36}$   \\
Core     &330 & $\rm 5.5\times 10^{35}$  &$\rm 3.5\times 10^{34}$  \\\hline
\end{tabular}
\end{table}


\subsection{Depolarization}\label{fractionl}

\begin{figure}
\resizebox{9.5cm}{7.5cm}{\includegraphics{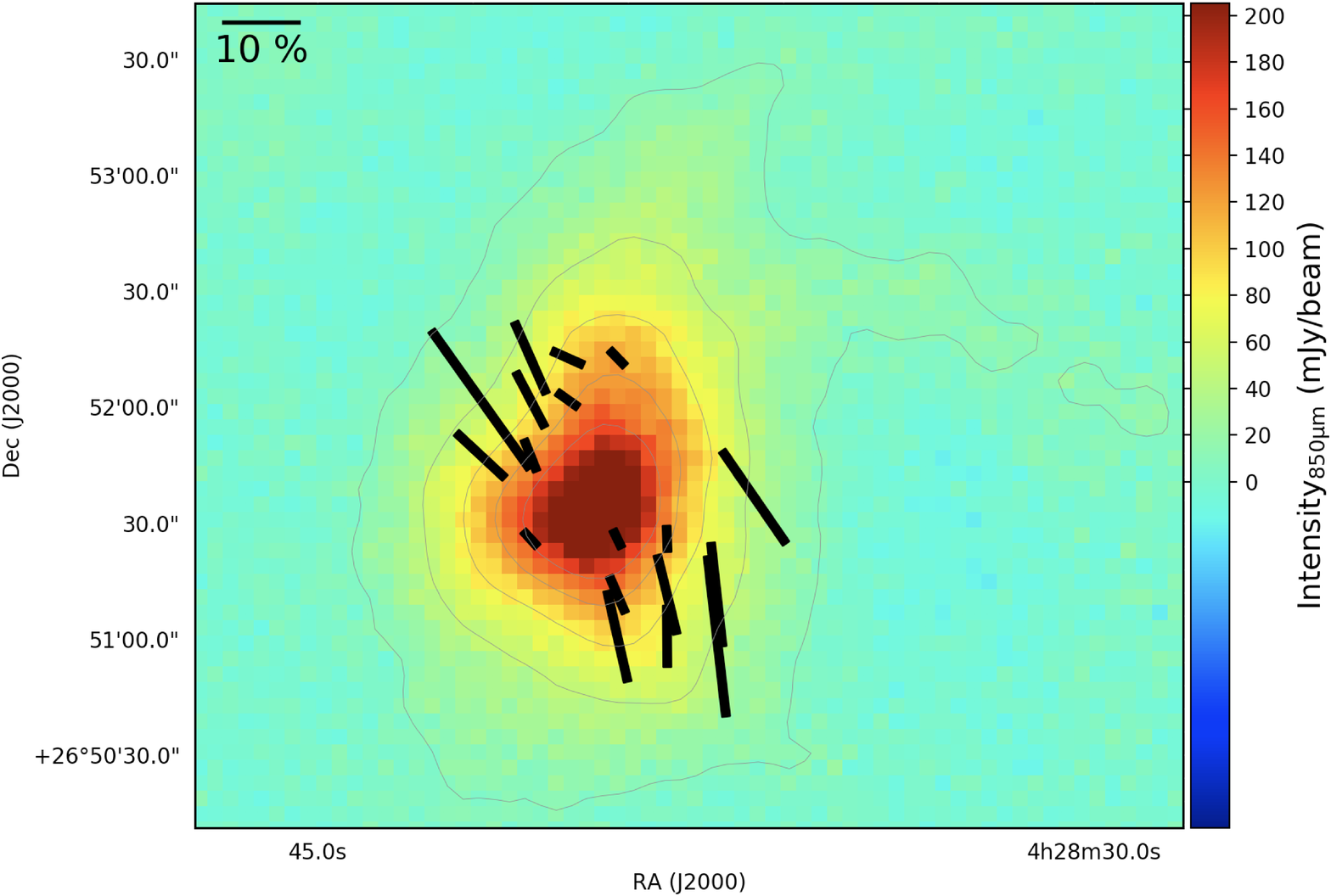}}\\
\resizebox{9.5cm}{8cm}{\includegraphics{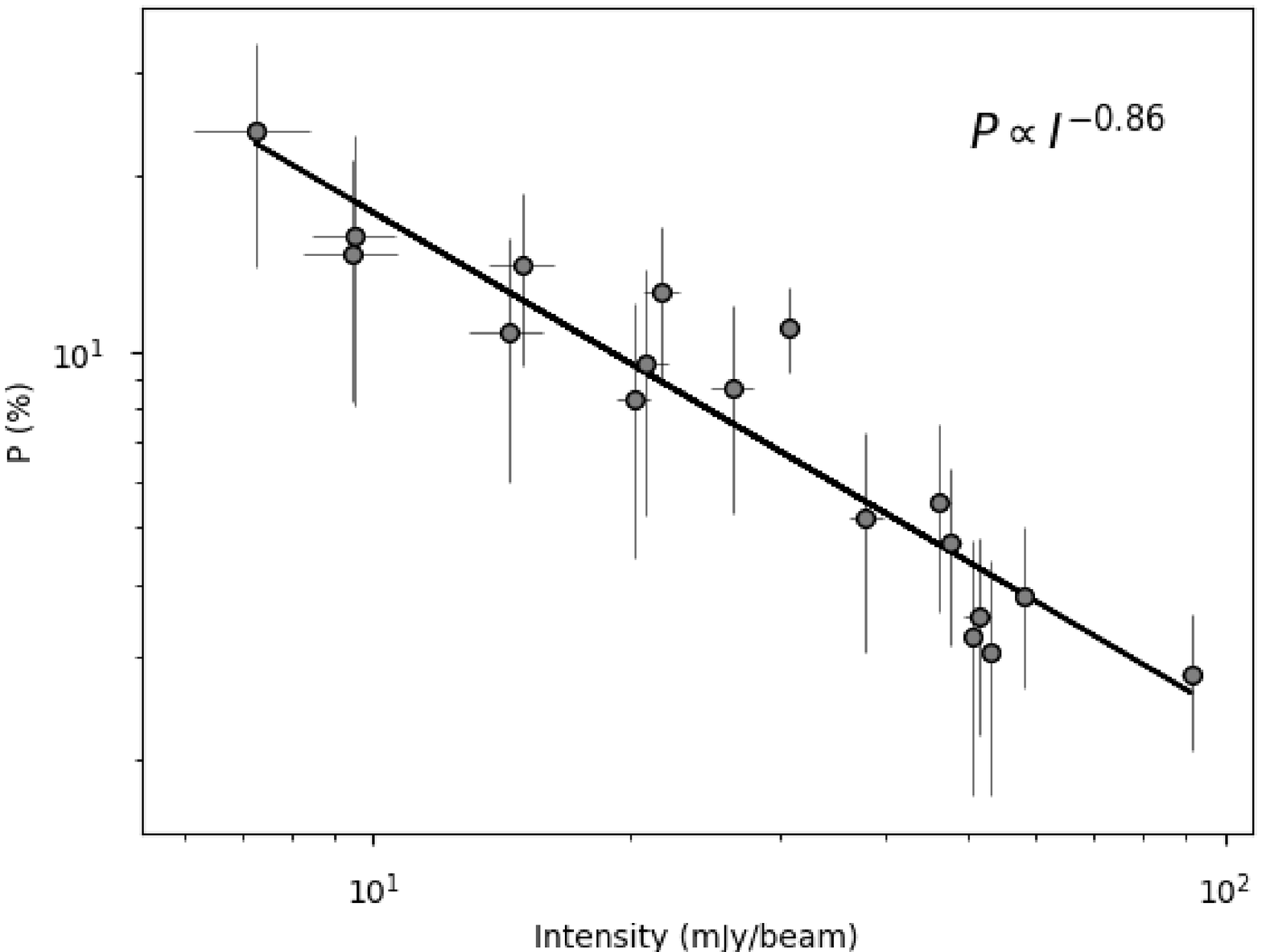}}
\caption{{\bf Upper panel:} The B-field geometry observed from 850$~\mu$m POL-2 observations. The lengths of line-segments are proportional to the polarization percentage. The scale for polarization fraction is also indicated. {\bf Lower panel:} The variation of polarization fraction with the intensity in L1521F core. The values and their uncertainties are taken from the POL-2 measurements with 12$\arcsec$ pixel size with a 3$\times$3 binning.} \label{Fig:pol_I}
\end{figure}

We investigated the change in polarization fraction from diffuse to high density regions on L1521F core as done in several other star-forming regions by previous studies \citep[i.e.][]{2000ApJ...531..868M, 2001ApJ...561..864L, 2013ApJ...763..135T, 2014A&A...569L...1A, 2017ApJ...847...92H, 2018ApJ...855...39K, 2018A&A...620A..26J}. Top panel of figure \ref{Fig:pol_I} shows the B-field geometry seen in L1521F core from POL-2 observations. The lengths of line-segments are proportional to the polarization percentage. It can be noticed that amount of polarization fraction decreases towards high-density brighter regions. The decrement can be quantified by comparing the length of line-segments with scale bar. This phenomena is widely known as \textit{depolarization} in molecular cloud cores. This variation of degree of polarization is further investigated in lower panel of the figure which shows the variation of polarization fraction with intensity. This plot suggests that the core-scale fraction of polarization is decreasing with the intensity towards the higher density regions. We measured the power-law slope in the distribution using a least-squares fit, and found it to be $\rm \alpha = -0.86$. The widely accepted possible reasons of lower observed polarization fraction in the cores include the changes in B-field orientation in the denser regions. These changes could also stem from the grain growth with a consequence of more spherical shapes thus not getting aligned with the B-fields. Other possible reasons of the change are suspected to be the magnetic reconnection \citep{1999ApJ...517..700L} and/or from weak Radiative Alignment Torques (RATs; \citealt{2007MNRAS.378..910L}) due to weak radiation field in dense regions. But the radiation from L1521F-IRS may compensate for the loss of  interstellar radiation field and help in radiative alignment of the grains.

The power-law index in L1521F is found to be similar to the index $\alpha \sim$ -0.9 in $\rho$ Ophiuchus B \citep{2018ApJ...861...65S} and more identical to  $\alpha \sim$ -0.8 in $\rho$ Ophiuchus A \citep{2018ApJ...859....4K}. A recent study by \citet{2018ApJ...855...81S} explored the reasons of the observed depolarization in the center of molecular clouds using dust polarization maps of 3D magnetohydrodynamical (MHD) simulations. From the MHD modeling of their synthetic data, they found that  dust grains remain aligned at even at high densities ($\rm > 10^{3} cm^{-3}$) and visual extinction of $\rm A_{V} > 1$ mag. They suggest that \textit{depolarization} is rather caused by strong variations of the magnetic field direction along the line-of-sight due to turbulent motions. 

\citet{1989ApJ...346..728J} and \citet{1992ApJ...389..602J} assumed that magnetic field consists of a constant as well as a random component and modeled polarized radiation through an extended cloud. They found $\rm \alpha = -0.5$ when random component dominates the magnetic field with expected polarization scales as the square root of number of turbulent cells. The values of slopes steeper than -0.5 leads to the prediction of more turbulent environment towards the central high density regions of the cores. We tested this hypothesis on our results by checking the slope in very center of the core. We considered the six data points with total intensities higher than 45\,$\rm mJy\,beam^{-1}$ from the lower panel of figure \ref{Fig:pol_I} for fitting power-law. We obtained a slope of -0.47 in central region which is much closer to the value found by \citet{1992ApJ...389..602J}. The shallower nature of the P vs I distribution in central high density region compared to the whole region suggests that the B-fields are stronger on the outer periphery and core is more turbulent towards central high density region causing scrambled B-fields and hence decrease in polarization due to beam averaging across lots of B-field orientations.

The turbulent nature of L1521F core is supported by several studies done to investigate the kinematics of this core using high resolution ALMA observations \citep{2014ApJ...789L...4T, 2016ApJ...826...26T, 2017ApJ...849..101T, 2018ApJ...862....8T}. The mentioned studies found that L1521F is a site with high-dynamic gas interactions and a multiple star formation at a scale of $\sim$100\,AU. \citet{2018ApJ...862....8T} observed $\rm ^{12}CO$(J=3-2) line in this core using ALMA at $\sim$20\,AU resolution and found a complex warm filamentary/clumpy structures with the sizes from a few tens of astronomical units to $\sim$1000 AU. These findings suggest that the plausible reason of depolarization seen towards L1521F may be the presence of enhanced turbulence in the central high density region of the core.

\subsection{Magnetic fields in other cores with VeLLOs}\label{other_cores}
The present study is the first sub-pc scale observation of a nearby core with VeLLO using high sensitivity POL-2 instrument as compared to the SCUPOL at JCMT. Prior to the investigation we presented in this work, there have been several attempts made for mapping B-fields in the low-mass star-forming cores. However, only one core with VeLLO namely IRAM 04191+1522 (hereafter IRAM04191) in Taurus has been observed with SCUPOL in a previous study by \citet{2009ApJS..182..143M}. We are considering this case with similar central source luminosity for the discussion in context with L1521F.

\citet{2015A&A...573A..34S} have mapped the B-fields in the envelopes of five cores with VeLLOs using optical polarization measurements. Polarization measurements towards IRAM04191 have been done in optical \citep{2015A&A...573A..34S} and 850 $\mu$m \citep{2009ApJS..182..143M} wavelengths. The mean values of the polarization fraction and the magnetic field orientation with corresponding standard deviations in IRAM04191 are found to be 13$\pm$7\% and 32$\pm$36$\degree$ using SCUPOL. It can be noticed from \citet{2015A&A...573A..34S} and this work, that B-field line changes their direction by almost 90$\degree$ in IRAM04191 whereas they got slightly bent in L1521F. This can be attributed to the different environments and kinematics of the two cores.

A highly collimated CO bipolar outflow was found to be associated with VeLLO embedded in IRAM04191 \citep{1999ApJ...513L..57A}. From the CO map of the region presented by \citet{2002A&A...393..927B}, the outflow direction is found to be 28$\degree$ in the plane of the sky from north towards east. The inner B-fields mapped with SCUPOL observations are found to be almost aligned with the outflows in IRAM04191. However, the outflows and core magnetic fields in L1521F are found to be misaligned with an offset of $\sim48\degree$.

The VeLLOs are supposed to preserve their inherent B-fields in the parent cores as they are sources at the lowest end of mass spectrum and the force from their associated bipolar outflows is found to be much smaller compared to other class 0 sources \citep{2019ApJS..240...18K}. This investigation can be done by plotting the variation of offsets between B-field orientation and outflows direction in these sources. \citet{2017MNRAS.464.2403S} has studied the detailed distribution of offsets between B-fields (inferred from optical polarization) and outflows with their corresponding outflow force values in several VeLLOs (shown in their figure 10). The outflows are found to be misaligned ($\sim$48$\degree$) in L1521F but aligned  ($\sim$4$\degree$) in IRAM04191. We need to increase the sample of such studies to arrive at any statistically significant conclusion of relation between outflow directions and magnetic field orientations.

\section{Summary}
\begin{enumerate}
\item We present results from the first sub-pc scale continuum polarization observation of L1521F, a core with an embedded VeLLO, now identified as a proto-brown dwarf candidate.  The B-fields are found very well connected to the large-scale field structures seen in Planck dust polarization and optical polarization measurements suggesting the core embedded in strong magnetic field region. The pronounced large scale field lines seems to be running in north-east and south-west direction.

\item The 850 $\mu$m continuum map interestingly shows two diffused elongated structures coming out of the main L1521F core.

\item The inner B-fields mapped with SCUPOL observations are found to be almost aligned with the outflows in IRAM04191. However, the outflows and core magnetic fields in L1521F are found to be misaligned with an offset of $\sim48\degree$.

\item The  B-field strength in L1521F core is estimated to be $\rm \sim330\pm100~\mu$G which is more than ten times larger than the value estimated in the envelope. The core is found to be super-critical with a $\rm \lambda$ value of 2.3$\pm$0.7.

\item The magnetic energies in the envelope and core are 1-2 orders of magnitude higher than the non-thermal kinetic energies of these regions making magnetic fields contributing more to the energy budget of L1521F.

\item The fraction of polarization as a function of total intensity is found to be decreasing in the denser region suggesting \textit{depolarization} in the core with a power-law slope of $\rm \alpha = -0.86$.

\end{enumerate}

The authors thank referee for the constructive comments and suggestions which helped in improving the content of the manuscript. AS and B-GA are supported by NSF Grant-1715876. AS carried out this work in part at the Korea Astronomy \& Space Science Institute (KASI) with support from the KASI postdoctoral fellowship. TL is supported by KASI and EACOA fellowships. MJ acknowledges the support of Academy of Finland grant 1285769. CWL is supported by Basic Science Research Program through the National Research Foundation of Korea (NRF) funded by the Ministry of Education, Science and Technology (NRF-2019R1A2C1010851). WK was supported by Basic Science Research Program through the National Research Foundation of Korea (NRF-2016R1C1B2013642). AS thanks P. Bhardwaj and Simon Coud{\'e} for the discussion during the analysis work. JCMT is operated by the East Asian Observatory on behalf of National Astronomical Observatory of Japan; Academia Sinica Institute of Astronomy and Astrophysics; the Korea Astronomy and Space Science Institute; the Operation, Maintenance and Upgrading Fund for Astronomical Telescopes and Facility Instruments, budgeted from the Ministry of Finance of China and administrated by the Chinese Academy of Sciences and, the National Key R\&D Program of China (No. 2017YFA0402700). 

\textit{Facility:} James Clerk Maxwell telescope (JCMT)
\textit{Softwares:} Starlink \citep{2014ASPC..485..391C}, Astropy \citep{2013A&A...558A..33A}.
\bibliographystyle{aasjournal}

\end{document}